\documentclass[amsmath,amssymb, aps, prl, twocolumn, superscriptaddress]{revtex4-2}
\usepackage{float}
\usepackage{graphicx}
\usepackage{dcolumn}
\usepackage{bm}
\usepackage{hyperref}
\hypersetup{
    colorlinks=true,
    linkcolor=blue,
    filecolor=magenta,      
    urlcolor=cyan,
}
\usepackage{color}
\usepackage{fancyhdr}
\usepackage[normalem]{ulem}

\begin{document}

\title{Observation of Fermi acceleration with cold atoms}

\author{G.~Barontini}
\email{g.barontini@bham.ac.uk}
\affiliation{School Of Physics and Astronomy, University of Birmingham, Edgbaston, Birmingham, B15 2TT, UK.}
\author{V.~Naniyil}
\affiliation{School Of Physics and Astronomy, University of Birmingham, Edgbaston, Birmingham, B15 2TT, UK.}
\author{J.~P.~Stinton}
\affiliation{School Of Physics and Astronomy, University of Birmingham, Edgbaston, Birmingham, B15 2TT, UK.}
\author{D.~G.~Reid}
\affiliation{School Of Physics and Astronomy, University of Birmingham, Edgbaston, Birmingham, B15 2TT, UK.}
\author{J.~M.~F.~Gunn}
\affiliation{School Of Physics and Astronomy, University of Birmingham, Edgbaston, Birmingham, B15 2TT, UK.}
\author{H.~M.~Price}
\affiliation{School Of Physics and Astronomy, University of Birmingham, Edgbaston, Birmingham, B15 2TT, UK.}
\author{A.~B.~Deb}
\affiliation{School Of Physics and Astronomy, University of Birmingham, Edgbaston, Birmingham, B15 2TT, UK.}
\author{D.~Caprioli}
\affiliation{Department of Astronomy and Astrophysics \& E. Fermi Institute, The University of Chicago, 5640 S Ellis Ave, Chicago, IL 60637, USA.}
\author{V.~Guarrera}
\affiliation{School Of Physics and Astronomy, University of Birmingham, Edgbaston, Birmingham, B15 2TT, UK.}

\date{\today}

\begin{abstract}
Cosmic rays are deemed to be generated by a process known as \emph{Fermi acceleration}, in which charged particles scatter against magnetic fluctuations in astrophysical plasmas. The process itself is however universal, has both classical and quantum formulations, and is at the basis of dynamical systems with interesting mathematical properties, such as the celebrated Fermi-Ulam model. Despite its effectiveness in accelerating particles, Fermi acceleration has so far eluded unambiguous verifications in laboratory settings. Here, we realize a fully controllable Fermi accelerator by colliding ultracold atoms against engineered movable potential barriers. We demonstrate that our Fermi accelerator, which is only 100 $\mu$m in size, can produce ultracold atomic jets with velocities above half a meter per second. Adding dissipation, we also experimentally test  Bell’s general argument for the ensuing energy spectra, which is at the basis of any model of cosmic ray acceleration. On the one hand, our work effectively opens the window to the use of cold atoms to study phenomena relevant for high energy astrophysics. On the other, the performance of our Fermi accelerator is competitive with those of best-in-class accelerating methods used in quantum technology and quantum colliders, but with substantially simpler implementation and virtually no upper limit.
\end{abstract}

\maketitle

Particle acceleration plays a central role in modern physics, for example in the high-energy colliders that have unlocked our understanding of the nuclear and sub-nuclear world, in the stability of fusion plasmas, and in the origin of extraterrestrial energetic particles. 
The energization of charged particles generally involves electromagnetic fields, with the accelerated particles revealing themselves through the non-thermal light that they emit, such as synchrotron radiation and inverse-Compton scattering. 
In astrophysical settings the energizing electromagnetic fields are typically of motional nature, given the impossibility of sustaining large-scale potential drops in tenuous plasmas. 
The prototypical example of extraterrestrial energetic particles is the cosmic ray radiation \cite{Longair_2011, gaisser}: ultra-relativistic ions and electrons produced in cosmic explosions and compact stellar remnants (novae, supernovae, pulsars, gamma-ray bursts), and relativistic jets launched by black holes (microquasars, active galactic nuclei).

The cornerstone mechanism for the origin of cosmic rays was proposed by E. Fermi \cite{fermi49, fermi54}: 
when a particle undergoes a head-on (tail-on) elastic collision with a moving magnetic mirror it gains (loses) kinetic energy, but since head-on encounters are more likely than tail-on ones (think about inverse-Compton scattering), on average the particle's energy must increase with time. {While in astrophysical plasmas this \textit{Fermi acceleration} is realized by scattering charged particles against magnetic fluctuations, the process itself is relevant for a wide range of terrestrial classical \cite{lieberman1998fermi,Sacha_2018,llPhysRevLett.74.4972} and quantum \cite{jose,pizzi2021higher, barbosa2024stabilizingultracoldfermigas} systems, and has inspired several mathematical models with unique properties \cite{ulam,brahic,lieberman}.} 
In the past two decades, kinetic simulations of astrophysical plasmas exploiting the growing power of modern supercomputers for solving Maxwell's equations coupled with a phase-space description of particles, have been able to shed light on how acceleration may occur in  astrophysical shocks \cite{malkov+01,spitkovsky08, karimabadi+14, caprioli+14a, caprioli+20}, in magnetic reconnection {\cite{druryx, sironi+11, cerutti+14, petropoulou+16}}, and in turbulence \cite{comisso+18,pecora+18,zhdankin+19}.
Experimentally, several impressive endeavors are working towards producing particle acceleration at shock in laboratory settings \cite{bbfiuza2020electron,bbPhysRevLett.122.245001,bbPhysRevLett.123.055002,bbyao2021laboratory,bbyuan2024electron}. However, to date only hints of deviations from thermal distributions have been reported.
In fact, sustaining a plasma laser experiment for many acceleration cycles to detect extended spectra of accelerated particles is still out of reach.

\begin{figure} 
	\centering
	\includegraphics[width=0.48\textwidth]{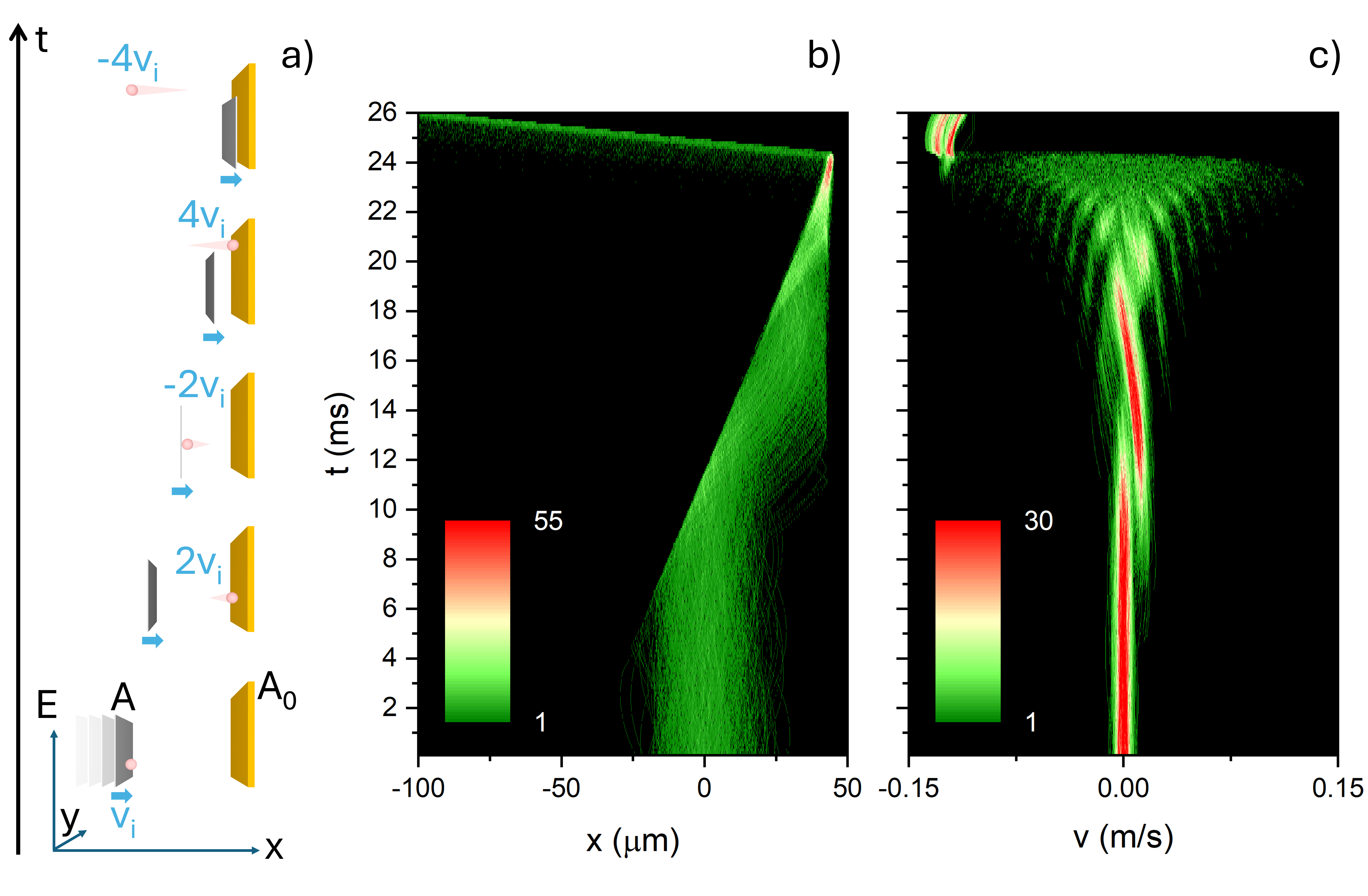}
	\caption{
		(\textbf{a}) A particle is hit multiple times by a potential barrier with height $A$ and moving with velocity $v_i$ along the $\hat{x}$ direction, and a stationary potential barrier with height $A_0\geq A$. Each collision with the moving barrier increases the particle's velocity by $2v_i$. When the kinetic energy of the particle is higher than $A$, the particle escapes the accelerator. (\textbf{b}) Numerically simulated time evolution of the density distribution along $\hat{x}$ of a cloud of $10^3$ non-interacting $^{87}$Rb atoms with $T=50$ nK, inside a Fermi accelerator with $A/k_B=0.7A_0/k_B=50$ $\mu$K and $v_i=3.64$ mm/s \cite{SM}. The atoms are also subject to a harmonic potential with 40 Hz trapping frequency \cite{SM}. The color scale indicates the density in $10^7$m$^{-1}$. The stationary barrier is placed at $x=45$ $\mu$m, while the moving barrier starts at $x=-45$ $\mu$m. The two barriers collide for $t=$25 ms. (\textbf{c}) The same as panel (b) but in the velocity space. The color scale indicates the density in $2\times10^4$(m/s)$^{-1}$.}
	\label{fig1} 
\end{figure}

In this work we realize a table-top Fermi accelerator in which neutral particles are accelerated with controlled potential mirrors. The basic principle in one dimension ($\hat{x}$) is shown in Fig. 1a), where we consider a particle initially at rest being  scattered between two potential barriers of height $A$ and $A_0\geq A$. 
If the barrier $A$ (left in the figure) moves with velocity $v_i$ and if the collision is elastic  in the barrier frame, after the collision the particle acquires a velocity $+2v_i$ in the laboratory frame.
The collision with the barrier $A_0$, instead, simply reverses its motion $v\to -2v_i$.
While traveling back, the particle is hit a second time by the moving barrier, increasing again its velocity by $2v_i$. The process repeats with the frequency of collisions rapidly increasing due to the increasing particle velocity and the decreasing distance between the barriers. When the kinetic energy of the particle $E_k=mv^2/2=m(2nv_i)^2/2$, with $m$ the mass of the particle and $n$ the number of collisions with the moving barrier, exceeds $A$, the particle leaves the accelerator. The net result is an acceleration of the particle to a final velocity of $-2nv_i$. 

As shown in Fig. 1b), the same mechanism holds also for a cloud of trapped ultracold atoms that are initially distributed in real and momentum space according to their temperature. The atomic cloud is compressed in between the two colliding barriers and, in absence of interactions, each atom undergoes the Fermi acceleration process, resulting in an atomic jet exiting the accelerator. The dynamics inside the accelerator can be better appreciated by looking at the characteristic fractal-like evolution in the velocity (or momentum) space, shown in Fig. 1c). Each collision with the barriers adds a `fringe' to the velocity distribution, increasingly broadening it. As the barriers come close together, the rate of collisions with the barriers drastically increases, and with it the broadening of the velocity distribution. This process continues until the threshold $v=\sqrt{2A/m}$ is reached and the atoms can leave the accelerator. Because this accelerator is based on a threshold process, the width of the distribution of the ejected atoms is essentially dictated by the width of the distribution at $t=0$ \cite{SM}.    

\begin{figure*} 
	\centering
	\includegraphics[width=0.6\textwidth]{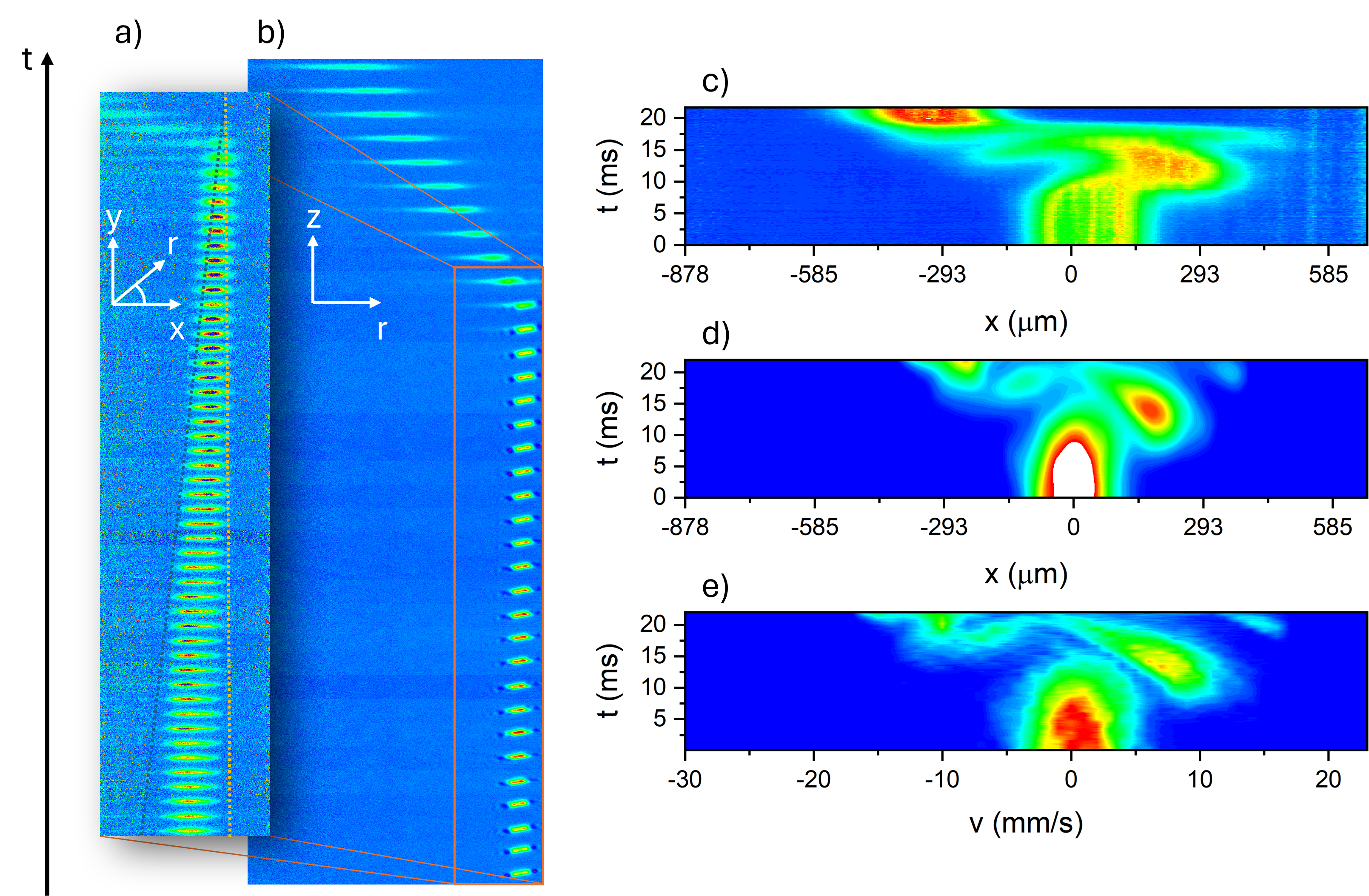}
	\caption{
		(\textbf{a}) Stacked in situ absorption images of the atomic cloud taken along $\hat{z}$. Snapshots are taken every 0.5 ms, the total time is 25.5 ms. Each snapshot is 243 $\mu$m along $\hat{x}$ and 64.5 $\mu$m along $\hat{y}$. The black and yellow dashed lines approximately indicate the position of the moving and stationary barrier respectively. At $t=0$ there are $\simeq25\times10^3$ atoms at $\simeq20$ nK. The moving barrier has a speed of $3.6$ mm/s, and $A/k_B=0.7A_0/k_B=170$ $\mu$K. (\textbf{b}) Stacked in situ absorption images of the atomic cloud taken along a direction that is tilted by 50 degrees with respect to $\hat{x}$. The experimental parameters are the same as in panel (a), but this imaging features lower resolution and larger field of view. Snapshots are taken every ms, the total time is 37 ms. Each snapshot is 1093 $\mu$m along $\hat{x}$ and 132 $\mu$m along $\hat{z}$. (\textbf{c}) The contour plot is the result of 250 stacked absorption images taken along the same direction as panel (b), and integrated along $\hat{z}$. Each absorption image has been taken after 22 ms of `time of flight' inside the waveguide oriented along $\hat{x}$. The horizontal scale has been rescaled to account for the 50 degrees angle between the imaging direction and  $\hat{x}$. Note that the images at low $t$ are heavily saturated. The parameters for the Fermi accelerator are $A/k_B=0.7A_0/k_B=1.8$ $\mu$K and $v_i=3.6$ mm/s. (\textbf{d}) Density distribution resulting from the numerical simulations with the same parameters of the experiment shown in (c) \cite{SM}. The color scale has been saturated to mimic the experiment. (\textbf{e}) Density distribution in velocity (momentum) space inside the accelerator for the simulations in (d). }
	\label{fig2x} 
\end{figure*}

Our Fermi accelerator is implemented with ultracold $^{87}$Rb atoms and potential barriers generated by laser light. {This latter indeed produces an optical dipole force \cite{grimm} whose strength can be controlled with the power and detuning of the laser beam \cite{SM}. The barriers are dynamically generated by using a digital micromirror device that shapes the profile of the laser beam \cite{SM}}. The atoms are initially trapped in a crossed beam dipole trap, which is deformed into a one dimensional waveguide along the direction of the accelerator before the acceleration process begins. This is done to support the atomic jet against gravity \cite{SM}. In Fig. 2a) we report a typical experimental sequence for an accelerator with $A/k_B=0.7A_0/k_B\simeq170$ $\mu$K, where $k_B$ is the Boltzmann constant, observed with absorption imaging. The moving barrier, whose position is approximately indicated by the black dashed line, hits the ultracold sample that is initially at rest. The center of mass of the atomic cloud then moves with velocity $2v_i$ towards the static barrier, indicated by the yellow dashed line. Similarly to Fig. 1b), the subsequent microscopic dynamics cannot be resolved in situ, due to the spread of the cloud in momentum and real space. Nonetheless, the net effect of the Fermi acceleration acting on all the atoms manifests itself with the atoms being ejected from the accelerator. As a result of the acceleration process, we observe atomic jets traveling at high speed in the waveguide, as shown in Fig. 2b). {Note that the direction of the atomic jet can be controlled with the $A/A_0$ ratio. For $A>A_0$, the atoms escape in the opposite direction (upstream) \cite{SM}}. 

\begin{figure} 
	\centering
	\includegraphics[width=0.48\textwidth]{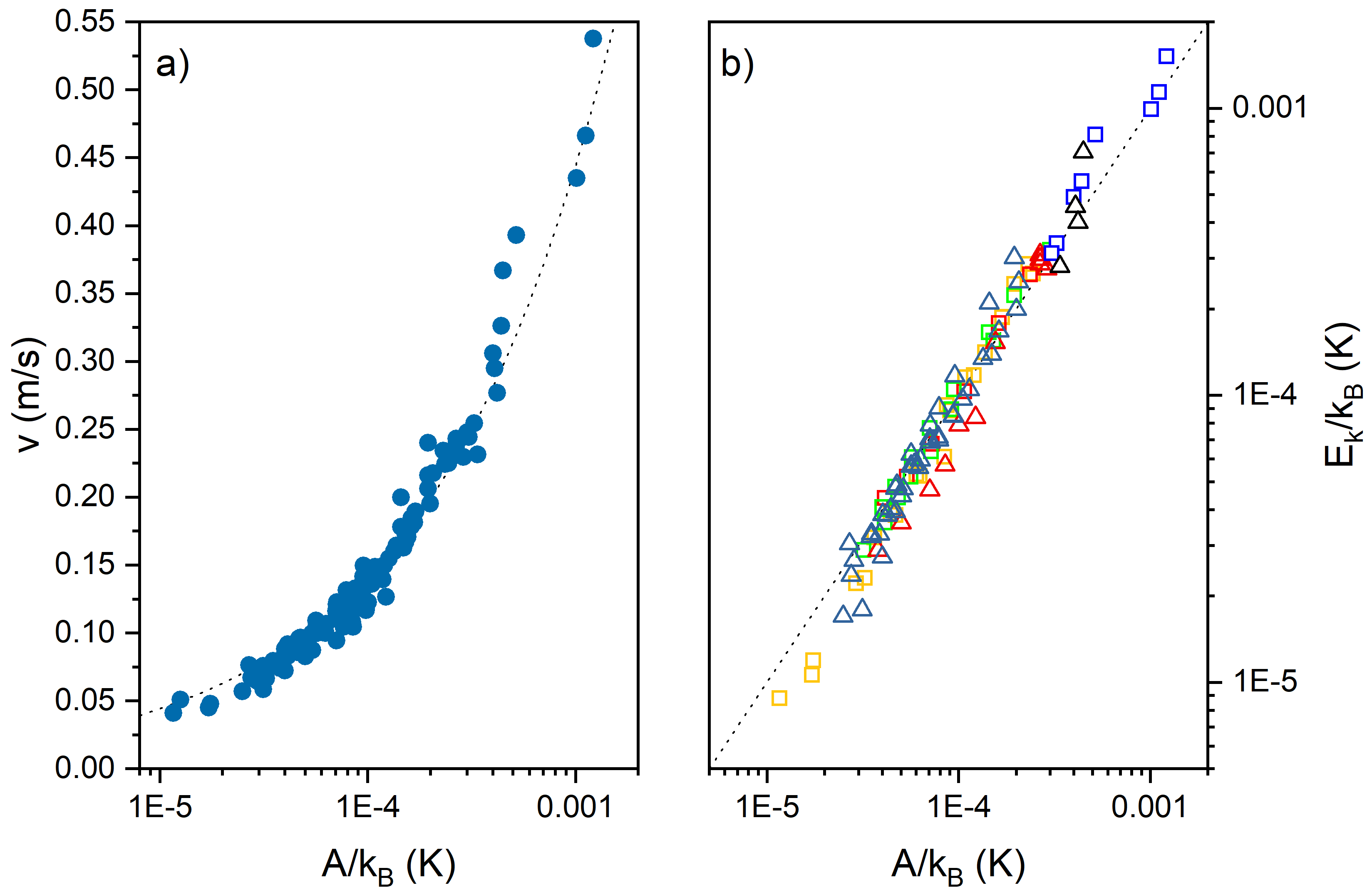}
	\caption{
		(\textbf{a}) Measured velocity of the atoms escaping the accelerator as a function of the height of the moving barrier $A$. The dotted line corresponds to $\sqrt{2A/m}$. (\textbf{b}) Kinetic energy of the particles exiting the accelerator for the data points reported in panel (a). Square symbols are for flat moving potential barriers, triangular symbols are for convex triangular moving barriers \cite{SM}. The parameters for the yellow squares are $v_i=11$ mm/s and $T$=200 nK, blue squares $v_i=22$ mm/s and $T$=200 nK, red squares $v_i=5.5$ mm/s and $T$=20 nK, green squares $v_i=4.4$ mm/s and $T$=20 nK. The parameters for the red triangles are $v_i=5.5$ mm/s and $T$=20 nK, black triangles $v_i=5.5$ mm/s and $T$=250 nK, blue triangles $v_i=4.4$ mm/s and $T$=20 nK \cite{SM}. Note that for the rightmost points the acceleration experienced by the atoms is $\simeq$10 times the gravitational acceleration. The dashed line is the bisector. The vertical errorbars are smaller than the size of the data points in both panels. The values of $A$ have a 5\% systematic uncertainty (not shown) coming from slow drifts in the focussing of the barriers. }
	\label{fig2} 
\end{figure}

{In our experiment, we can observe the full evolution in momentum space by utilizing barriers with relatively low values of $A$.} As an example, in Fig. 2c) we report the measured density distribution along $x$ after 22 ms of time-of-flight, for an accelerator that has a rather low moving barrier with $A/k_B\simeq1.8$ $\mu$K. We can clearly observe two collisions with the moving barrier (the two fringes on the right) and two with the stationary barrier (fringes on the left). After the second collision with the fixed barrier, the atoms have sufficient energy to fly over the moving barrier and exit the accelerator. In Fig. 2d) we report the evolution according to our numerical simulations for the same parameters as Fig. 2c) \cite{SM}, finding remarkable agreement. Fig. 2e) displays the corresponding density distribution in the velocity space that, as expected, mirrors what is observed in time-of-flight. Note the spreading of the momentum distribution while the atoms are inside the accelerator, and the narrow momentum distribution of the atomic jet. {For larger values of $A$, the observation of the full evolution of the density distribution in momentum space becomes challenging. In fact, as the spread in velocity increases, we cannot use standard time-of-flight methods, since most of the atoms rapidly fly out of our field of view.}

We infer the velocity of the atomic jet by measuring the position of its center of mass as a function of time \cite{SM}. The results for a wide range of values of $A$ are reported in Fig. 3a). By controlling $A$ over two orders of magnitude, we can finely control the speed of the atomic jet. Note that, despite the fact that the maximum speed of our barrier is limited to $22$ mm/s \cite{SM}, we are able to accelerate our ultracold cloud up to more than 0.5 m/s {in 5 ms}. This is competitive with what is obtained in free-fall with sequential Bragg large momentum transfer techniques \cite{kasevich} and Bloch oscillations in an optical lattice \cite{guellati-khelifa}, and more effective than other in-trap techniques based, for example, on magnetic time-averaged adiabatic potentials \cite{klitzing}. As discussed in \cite{SM}, the maximum velocity achievable in our experiment is limited only by atom losses caused by the use of near-resonant light for the generation of the optical potentials. There are however no fundamental limitations to the maximum velocity achievable, if dissipation is minimized \cite{SM}. In Fig. 3b) we report the calculated kinetic energy for the data of Fig. 3a), where it is apparent that $E_k\simeq A$ independently on the shape and velocity of the barriers, and the temperature of the atoms (see also \cite{SM}). 

From Fig. 2b) one can observe that the width of the accelerated cloud increases as the atoms propagate in the waveguide. The speed at which the atomic jet spreads is given by the width of the distribution in momentum space, and it can be linked to an `effective temperature' \cite{SM}. The comparison of our data with a simple numerical model shows that the effective temperature in our experiment, which is typically between 1 and 10 $\mu$K depending on the details of the accelerator \cite{SM}, is mainly determined by atom-atom collisions during the acceleration \cite{SM}. In fact, as the barriers approach each other the density of the sample and the average velocity greatly increase, leading to a boost of the atom-atom collisional rate. Such collisions redistribute part of the gained kinetic energy in the perpendicular directions, and lead to a `dephasing' of the Fermi mechanism. Remarkably, we have found however that our Fermi accelerator is robust against interactions, as they have no impact on the velocity of the atomic jet (Fig. 3) \cite{SM}, and {their effect} on the effective temperature can be drastically reduced or controlled by engineering the shape of the barriers \cite{SM}. 

\begin{figure} 
	\centering
	\includegraphics[width=0.48\textwidth]{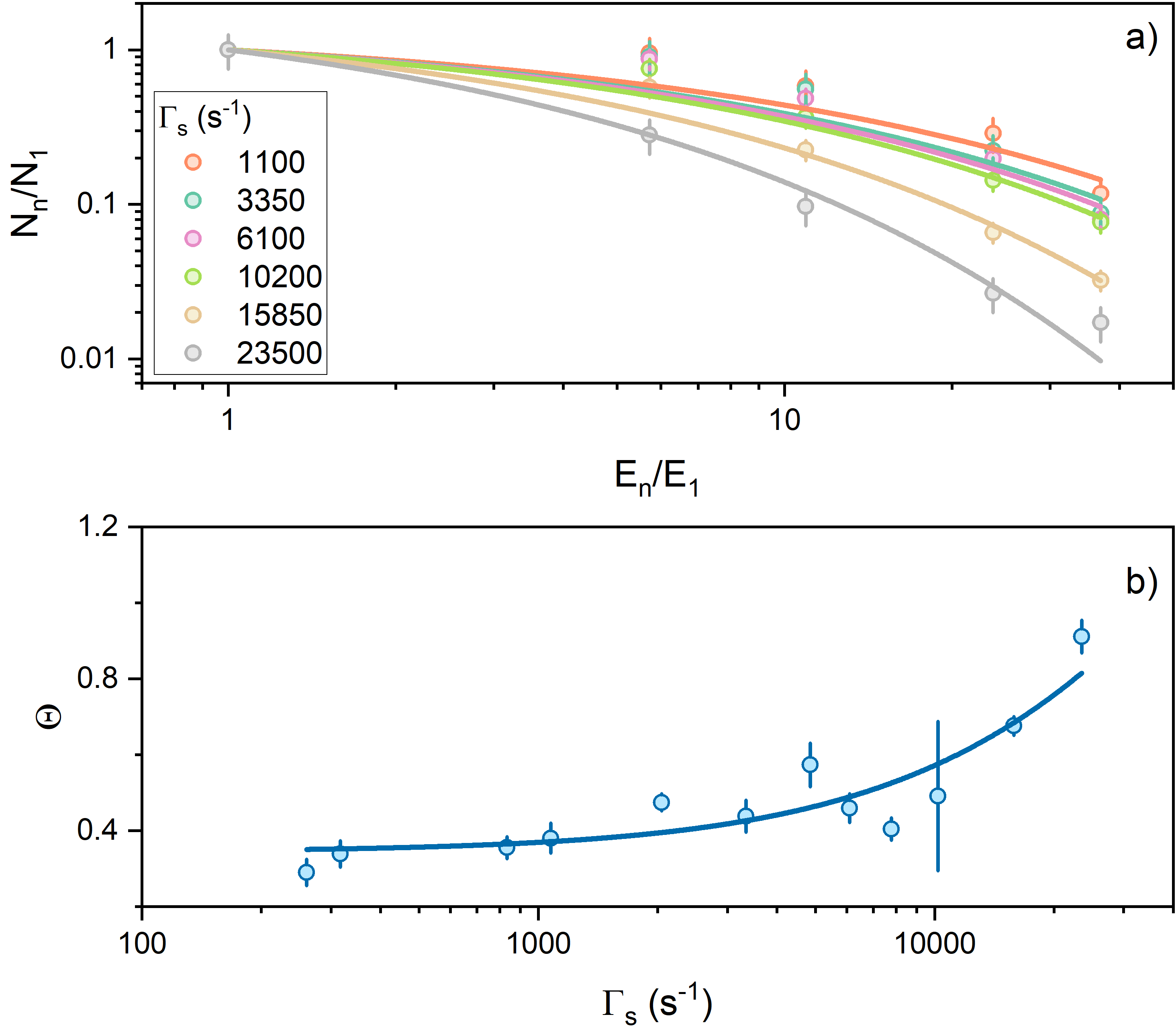}
	\caption{
		(\textbf{a}) The solid symbols are the number of atoms remaining inside the accelerator as a function of their average kinetic energy, for different values of the light-atom scattering rate.  Each data point is taken after a collision with the moving barrier. Both the number of atoms and the energy are normalized with respect to the figures we obtain after the first collision with the barriers \cite{SM}. The errorbars are 1 $\sigma$ statistical errors. The solid lines are fits to the data with a power-law function \cite{SM}. 
        (\textbf{b}) Data points are the value of the exponent $\Theta$ obtained by the fits in panel (a) and additional data shown in \cite{SM}, as a function of the calculated scattering rate stemming from the potential barriers. The errorbars come from the fitting procedure. The solid line is a fit to the data with the function $\Theta=\ln(\gamma_0+\xi\Gamma_s)$, with $\gamma_0=(2.1\pm0.2)$ and $\xi=(1.1\pm0.2)\times10^{-4}$ s.} 
	\label{fig4} 
\end{figure}

We finally utilize our setup to experimentally test the famous Bell's argument \cite{Bell}, which is one of the cornerstones of modern high-energy astrophysics, where extended power-law distributions are routinely observed, both in the cosmic radiation (electrons, nuclei, and neutrinos) and in the non-thermal emission from astrophysical objects. 
Bell argued that power-law distributions generally arise as the result of a combination of energy gain and particle escape from the accelerator. 
In our system, Bell's argument predicts that the energy spectrum inside the accelerator must have the form $N_n=N_1(E_n/E_1)^{-Q}$ with $Q=-\ln(P)/\ln(G)$, where $N_1 (E_1)$ and $N_n (E_n)$ are the number of atoms (average energy) after one and $n$ collisions with the moving barrier respectively. $P$ and $G$ are the survival probability and energy gain after each collision \cite{SM}. To verify this prediction, we slightly modified the experimental sequence so to measure the number of atoms left and their average velocity after each collision with the barriers \cite{SM}. By controlling the detuning and power of the laser light that generates the potential barriers, we are able to tune the light-atom scattering rate $\Gamma_s$, which in turn determines $P$, leaving all the other parameters of the experiment unchanged. {For this set of data we work close to resonance to increase $\Gamma_s$ \cite{SM}.}
In Fig. 4a) we report our results on the number of atoms left in the accelerators as a function of their average energy, for different values of $\Gamma_s$. Our data are well fitted  by the power-law function described above (solid lines), with {$Q=\Theta/\ln(G)$}, where $\Theta$ is the fitting parameter and $G$ is a function of $E_n/E_1$ \cite{SM}. In general, atom losses in our system are a combination of losses coming from the light-atom scattering discussed above and other loss mechanisms that don't depend on the detuning, the most relevant being three-body collisions. 
For this reason, in our experiment we expect the survival probability to have the form $P=(\gamma_0+\xi\Gamma_s)^{-1}$, where $\gamma_0$ accounts for all the detuning-independent loss mechanisms and $\xi$ for the effective interaction time of the atoms with the barriers. As shown in Fig. 4b), {the scaling of the $\Theta$ parameter observed in our experiment is remarkably compatible with the functional dependence} {$\Theta=\ln(\gamma_0+\xi\Gamma_s)\equiv -\ln(P)$} (solid line), as predicted by Bell's argument.   

Our results show that cold atom technology is an ideal platform to {implement} and study Fermi acceleration processes and related phenomena. For example it can be used as a testbed for classical, dissipative and quantum Fermi-Ulam models \cite{ulam,brahic,lieberman,jose}. {While several astrophysical phenomena will probably remain out of reach for table-top experiments, by combining the many tools available in cold atom systems one could investigate some of the open questions concerning particle acceleration at shock \cite{krymskii1977regular,axford1977acceleration,blandford1978particle,Bell,bell1978acceleration}, but also emerging phenomenology linked to propagating Alfv\'en waves \cite{zuzzoPhysRevLett.132.215201}}, {for example using multiple moving barriers}. Our methods could be straightforwardly applied to ultracold neutral plasmas \cite{killian,simonet}, where some of the conditions found in astrophysical settings could be obtained \cite{ichimaru}. {Particularly interesting is the possibility of controlling the strength and the nature of the interactions \cite{feshbach}, which is a major challenge in many-body numerical simulations. The experiment could test how different kinds of interactions affect the acceleration rate and therefore the maximum energy attainable. In addition, one could compare diffusion and Levy flights regimes, study super-diffusion, and implement variations or periodicity in the speed of the scattering centers, which are all important aspects in astrophysical environments. The complete control over the potential barriers is another key element that could be exploited: for example one could dynamically change the height and the speed of the barriers to accelerate in different ways different classes of atoms and simulate the effect of a cosmic ray shock precursor. One possible extension of our experimental setup is the study of stochastic (or second-order) Fermi acceleration, which is potentially important in astrophysics \citep{mertsch+11, brunetti+14}. Differently from first-order acceleration in shocks, stochastic acceleration does not lead to universal power laws in energy, and the theory that connects the statistics of scattering agents and particle energization is an active area of research \cite{lemoine25}.
Using a superposition of potentials with a prescribed power spectrum and measuring the ensuing particle spectrum may inform models of cosmic ray acceleration in both galactic and extragalactic environments. } 

In the context of quantum technology, our Fermi accelerator enables to obtain samples with large $E_k$ and low effective temperatures, that are desirable features in quantum chemistry \cite{softley,ye,narevicius} and quantum colliders applications \cite{legere,gibble,amita}. With respect to other record-achieving techniques in atom optics \cite{robins}, a Fermi accelerator is characterized by a remarkably simple implementation, which makes it ideally suited for in-trap atomtronics applications \cite{atomtronics}. Fermi acceleration can be also used to direct and split atomic samples into two counter-propagating clouds \cite{SM}, thus providing complementary features to existing techniques \cite{kasevich,guellati-khelifa}. Finally, the study of quantum Fermi acceleration \cite{jose, SM} could be a promising avenue for quantum information science, as there is concrete potential for the development of new tools for the manipulation of quantum wavepackets . 

\noindent
\textit{Acknowledgments:}
We acknowledge fruitful discussion with the cold atoms group and the theoretical physics group at the University of Birmingham, and the use of computing power provided by the Advanced Research Computing centre at the University of Birmingham.
This work was supported by EPSRC through grants EP/V027948/1, EP/R021236/1, 2740639 and EP/W016141/1, by NASA through grants 80NSSC18K1218, NSSC23K0088, and 80NSSC24K0173, by NSF through grants PHY-2010240 and AST-2009326, and by the Royal Society through grants UF160112, RGF/EA/180121 and RGF/R1/180071.

\bibliography{FA_bib} 

\end{document}